\documentclass[a4paper,11pt]{article}
\pdfoutput=1 

\usepackage{jheppub} 

\usepackage[T1]{fontenc} 
\usepackage{slashed}
\usepackage{graphicx,color}
\usepackage{dcolumn}
\usepackage{bm}
\usepackage{subfig}
\usepackage{graphicx}
\usepackage{amssymb}
\usepackage{stackrel}
\usepackage{pgfplots}
\usetikzlibrary{positioning}
\usetikzlibrary{snakes}
\newcommand{\be}{\begin{equation}}
\newcommand{\ee}{\end{equation}}
\newcommand{\bea}{\begin{eqnarray}}
\newcommand{\eea}{\end{eqnarray}}

\begin{document} 
\title{\boldmath Neutrino oscillation in dark matter with $L_\mu-L_\tau$}


\author[a]{Wei Chao,}
\author[b]{Yanyan Hu,}
\author[a]{Siyu Jiang,}
\author[a]{Mingjie Jin,}


\affiliation[a]{Center for advanced quantum studies, and Department of Physics,
Beijing Normal University, \\ Beijing 100875,  China}
\affiliation[b]{
Key Laboratory of Beam Technology of Ministry of Education,
College of Nuclear Science and Technology, Beijing Normal University, Beijing 100875, China }
\emailAdd{chaowei@bnu.edu.cn}
\emailAdd{huyy@mail.bnu.edu.cn}
\emailAdd{jiangsy@mail.bnu.edu.cn}
\emailAdd{jinmj@bnu.edu.cn}

\abstract{In this paper, we study the phenomenology of a Dirac dark matter in the $L_\mu-L_\tau$ model and investigate the neutrino oscillation behavior in the dark halo. Since dark matter couples to muon neutrino and tau neutrino with opposite sign couplings, it contributes effective potentials, $\pm A_\chi$, to the evolution equation of the neutrino flavor transition amplitude, which can be significant for high energy neutrino oscillations in a dense dark matter environment. We discuss neutrino masses, lepton mixing angles, Dirac CP phase, and neutrino oscillation probabilities in the dark halo using full numerical calculations. Results show that neutrinos can endure very different matter effects. When the potential $A_\chi$ becomes ultra-large,  three neutrino flavors decouple from each other.}

\maketitle
\flushbottom

\section{Introdution}
\label{sec:intro}

Robust evidences from astrophysical observations point to the existence of cold dark matter (DM), which can not be addressed by the minimal Standard Model (SM) of particle physics.  For a DM mass above ~$1~{\rm keV}$, it behaves as a cold DM~\cite{Steigman:1984ac}. It is well-known that cold DM needs to interact with the SM particles in additional to the gravitational interaction so as to explain the observed relic abundance, but how it couples to the SM is still unknown. There are various DM candidates with mass ranging from $10^{-22}~{\rm eV}$ to $10^{55}~{\rm GeV}$. Due to the progress of direct and indirect DM detection technology,  many DM models have already been excluded, however the neutrino portal~\cite{Dodelson:1993je,Adhikari:2016bei,Chao:2020bti,Bertuzzo:2018itn,Berlin:2018ztp,Becker:2018rve,Batell:2017cmf,Okada:2016tci,Escudero:2016ksa,Escudero:2016tzx,Cherry:2014xra} is relatively safe, as neutrinos themselves are also difficult to probe. One of the most famous neutrino portal model is sterile neutrino DM which can be warm or cold DM and whose production mechanism is via neutrino oscillations~~\cite{Dodelson:1993je,Adhikari:2016bei}. 
There is also famous scotogenic model~\cite{Farzan:2012sa}  which include the DM and neutrino masses into one framework by introducing extra Yukawa interactions.  A third typical neutrino portal model takes $Z^\prime$ or dark photon as the mediator~\cite{Li:2010rb,Cai:2014hka}.  New interactions in the neutrino portal may induce irreducible background, named as ``neutrino floor"~\cite{Monroe:2007xp,Strigari:2009bq,Billard:2013qya,Chao:2019pyh}, in direct detection experiments.  They may also generate some exotic signals in various neutrino oscillation experiments. It is of great significance to study these signals because they may be  an important indirect evidence for the existence of CDM.

In this paper, we will study  possible signal of DM in neutrino oscillations. As we all know, dark matter accounts for 26.8\% of the Universe, and the entire Milky Way Galaxy is in a huge dark halo. When neutrinos propagate in the dark halo, the interaction between neutrinos and DM will lead to the matter effect of neutrino oscillations. Note that the density of DM is about 0.4 ${\rm GeV / cm^3}$~\cite{Read:2014qva} near the solar system, so the matter effect caused by DM may be  too small to be observed in long baseline neutrino oscillation experiments that are located on the Earth. However, in some regions of the Universe, such as the center of the Galactic center or dwarf spheroidal galaxies, the density of DM can be very high. When high-energy neutrinos pass through these regions, strong matter effect can be induced by the DM. In fact, the matter effect induced by DM has drawn the theorists'  attention, and some important issues have already been addressed~\cite{Choi:2019zxy,Liao:2018byh,Capozzi:2018bps}.  
We will discuss the phenomenology of DM  and neutrino oscillations within the framework of $U(1)_{L_\mu-L_\tau}$~\cite{He:1991qd,Altmannshofer:2014cfa}, which is one of the most economical extensions to the SM. It does not require the introduction of additional elementary particles to eliminate various anomalies. Compared with $L_e-L_\mu$~\cite{Duan:2017qwj}, $B-L$~\cite{Langacker:2008yv}, and $U(1)_R^{}$~\cite{Chao:2017rwv}, this model is less restricted and can be used to explain exotic phenomena in high-energy physics experiments, such as the universality violation in the decay of B meson~\cite{Altmannshofer:2014cfa} and the low-energy recoil signal of XENON1T~\cite{Aprile:2020tmw}.  
%
%
We first perform a systematic study on  constraints on the model arising from the observed relic abundance of DM, upper limits on the direct detection cross section as well as the anomalous magnetic moment of the muon.  Then we   discuss impacts of this new neutral current interaction to neutrino oscillations. In the three-flavor neutrino oscillation scheme, we study the (dark) matter effect of  neutrino masses, lepton mixing angles, Dirac CP phase, and neutrino oscillation probabilities in the dark halo using full numerical calculations.  Our results are applicable to study high energy neutrino oscillations in a dense dark matter environment.

The remaining of the paper is organized as follows: In section II we introduce the model in detail and discuss various constraints. In section III, we study the phenomenology of DM in the $L_\mu-L_\tau$ model. Section IV is devoted to the study of neutrino oscillations in dark halo. The last part is concluding remarks.

\begin{table}
	\centering
	\begin{tabular}{c|c|c|c|c|c|c|c|c}
		\hline
		\hline
		Particles& $\ell_e $ & $\ell_\mu $ & $\ell_\tau $ & $e_R $ & $\mu_R $ & $\tau_R$ & $\chi_{L,R}^{} $ & $\Phi $\\ 
		\hline
		Charges & 0 & 1 & -1 & 0 & 1 & -1& 1 & 1\\
		\hline
		\hline
	\end{tabular}
\caption{Particles and relevant $U(1)_{L_\mu-L_\tau}$ charges.}
\label{tab0}
\end{table}

\section{The $U(1)_{L_\mu-L_\tau}$ model}
It is well-known that~\cite{He:1991qd} the differences of lepton numbers can be gauged U(1) symmetries with anomalies automatically cancelled.  Such gauge theories, named as $U(1)_{L_\alpha-L_\beta}$ with $\alpha,\beta=e,\mu,\tau$, have been widely studied as potential candidates of new physics beyond the SM. In this paper, we introduce a vector-like fermion $\chi$ in  the  $U(1)_{L_\mu-L_\tau}^{}$ model  to address the DM problem,  and study neutrino oscillations in the DM halo. The particles and relevant charge assignments  are shown in the table.~\ref{tab0}. The Lagrangian for new particles can be written as  
\begin{eqnarray}
{\cal L} \sim \overline{\chi } i\slashed{D} \chi + (D_\mu \Phi)^\dagger (D^\mu \Phi)- m_\chi\overline \chi \chi + \mu^2 \Phi^\dagger \Phi -\lambda (\Phi^\dagger \Phi)^2  
\end{eqnarray}
where $D_\mu =\partial_\mu^{} -i g_X^{} Z^{\prime}_\mu$ being the covariant  derivative with $g_X^{}$ the new gauge coupling and $Z^\prime $ the new gauge boson, $\Phi \equiv (\phi+i\eta)/\sqrt{2} $ being a complex scalar singlet. When $\Phi$ develops a non-zero vacuum expectation value (VEV) $v_\Phi^{}$, the $U(1)_{L_\mu-L_\tau} $ gauge symmetry is broken spontaneously and $Z^\prime $ gets nonzero mass $M_{Z^\prime} = g_X^{} v_\Phi$. The  physical scalar singlet is $\phi$ with the mass squared $M_\phi^2 =2 \lambda v_\Phi^2$. Here we have assumed that the mixing between $\Phi$ and the SM Higgs is negligible for simplicity. In the following we will address several  constraints that are relevant to  this model.    

\subsection{neutrino masses}
The discovery of neutrino oscillations have proved that the SM is incomplete and one needs to explain the origin of tiny but non-zero neutrino masses. A most economic approach towards understanding the origin of neutrino masses is using the dimension-five Weinberg operator~\cite{Weinberg:1979sa},
\begin{eqnarray}
{1\over 4} \kappa_{fg} \overline{\ell_{Lc}^{C}}^f\varepsilon_{cd} H_d \ell_{Lb}^g \varepsilon_{ba} H_a + {\rm h.c.} 
\end{eqnarray}
where $f$ and $g$ are flavor indices, $a,b,c$ and $d$ are isospin indices, $H$ is the SM Higgs doublet, $\ell_L$ is left-handed lepton doublet. This   operator comes from integrating out heavy seesaw particles. 
In the $U(1)_{L_\mu-L_\tau}$ model, $\ell_L^{e}$, $\ell_L^{\mu}$ and $\ell_L^{\tau}$ carry different $U(1)_{L_\mu-L_\tau}^{}$ charges. As a result, the active neutrino mass matrix takes the following form
\begin{eqnarray}
M_\nu  \sim \left(  \begin{matrix}\star & 0 & 0 \\ 0 &0 &  \star \\ 0 & \star & 0 \end{matrix} \right) \; ,
\end{eqnarray}
which results in $\theta_{12} =\theta_{13} =0$ and $\theta_{23}=45^o$  with $\theta_{ij}$ the mixing angle of the PMNS matrix in the standard parameterization. This scenario has been ruled out by the neutrino oscillation data.   One possible way out is including the following dimension-six effective operators 
\begin{eqnarray}
{1\over 4} \kappa_{e\mu}^\prime \Phi^\dagger \overline{\ell_{Lc}^{C}}^e\varepsilon_{cd} H_d \ell_{Lb}^\mu \varepsilon_{ba} H_a+{1\over 4} \kappa_{e\tau}^\prime \Phi \overline{\ell_{Lc}^{C}}^e\varepsilon_{cd} H_d \ell_{Lb}^\tau \varepsilon_{ba} H_a + {\rm h.c.} 
\end{eqnarray}
Then, only  the $(2,~2)$ and $(3,~3)$ elements in the neutrino mass matrix are zero. It has been shown in the Ref.~\cite{Meloni:2014yea} that this kind of texture zero only favors the inverted hierarchy scenario.  Actually  $(2,2)$ and $(3,3)$ elements can be nonzero by introducing dimension-seven effective operators. Taking into account these arbitrariness, we will not concentrate on the flavor structure of neutrino mass matrix in the $U(1)_{L_\mu-L_\tau}$ and take the experimental observables as input in the following study. It should be mentioned that these high dimensional operators may come from integrating out heavy right-handed Majorana neutrinos.    

\subsection{ muon g-2}

The anomalous magnetic moment of the muon, $(g-2)_\mu$ is one of the most precisely measured quantities in high energy physics.  Its experimental value is~\cite{Bennett:2006fi}
\begin{eqnarray}
a_\mu^{\rm exp} =116592089(63) \times 10^{-11} \; ,
\end{eqnarray} 
which deviates from the SM prediction~\cite{Keshavarzi:2019abf,Davier:2019can} by about $3.7\sigma$.  Due to the gauge interaction of muon with $Z^\prime$, $a_\mu$ receives contribution from the $Z^\prime$ mediated loop, which can be expressed as 
\begin{eqnarray}
\Delta a_\mu = {g_X^2 \over 4\pi^2} \int_0^1 dt {t^2(1-t) \over t^2 + (1-t){M_{Z^\prime}^2 /m_\mu^2}}
\end{eqnarray}  
As can be seen, both $g_X$ and $M_{Z^\prime}$ are relevant to the  $\Delta a_\mu$, which is always positive in this model. We show in the Fig.~\ref{fig:mchi_mzp} contours of $\Delta a_\mu$ in the $M_{Z^\prime} -g_X$ plane. The magenta band is favored by the current data. For more discussions about the $(g-2)_\mu$ in $U(1)_{L_\mu-L_\tau}$, we refer the reader to Refs~\cite{Heeck:2011wj,Gninenko:2018tlp,Amaral:2020tga} and references cited therein.

\subsection{ stability of the $Z^\prime$, $\phi$ and $\chi$ }

There are three new particles in the $L_\mu-L_\tau$ model: $Z^\prime$, $\phi$ and $\chi$. We will discuss their stabilities one by one. $Z^\prime$ couples to $\chi$, left-handed active neutrinos and charged leptons $\mu,\tau$. The total decay rate of $Z^\prime$ to leptons can be written as 
\begin{eqnarray}
\Gamma_{Z^\prime\to \bar f f}^{} &=&  \sum_{\alpha = \mu, \tau,\chi} \Theta(M_{Z^\prime} -2 m_\alpha )  {\alpha_X M_{Z^\prime} \over 3 } \sqrt{1 - 4 \beta_\alpha } \left( 1+ 2 \beta_\alpha  \right)+ {M_{Z^\prime } \alpha_X \over 3 } \label{zprate}
\end{eqnarray}
where $\alpha_X=g_X^2/4\pi$ and $\beta_\alpha =m_\alpha^2/M_{Z^\prime}^2$, $\Theta(x)$ is the step function. The second term on the right-handed side of the Eq. (\ref{zprate}) is the total decay rate to neutrinos in which we have neglected the tiny neutrino masses.   For $Z^\prime $ mass smaller than $2m_\nu$ where $m_\nu$ is the neutrino mass, the dominate decay channel is to three photons through the muon loop, and the decay rate is calculated as~\cite{Redondo:2008ec} 
\begin{eqnarray}
\Gamma_{Z^\prime \to 3 \gamma} \approx  {17 \alpha^3 \alpha_X \over 11664000 \pi^3} {M_{Z^\prime}^9 \over m_\mu^8}
\end{eqnarray}
where $\alpha $ is the fine-structure constant. In this paper, we assume that $M_{Z^\prime}$ is sizable and thus $Z^\prime$ cannot be a DM candidate. 

The physical scalar $\phi$ arises from the spontaneous breaking of the $U(1)_{L_\mu-L_\tau}$. It only couples to $Z^\prime$ in the toy model. For  $M_{\phi} > M_{Z^\prime}$, $\phi$ can decay into $Z^\prime$ pair, with the decay rate
\begin{eqnarray}
\Gamma_{\phi\to Z^\prime Z^\prime} = \alpha_X^{} {M_{Z^\prime}^2\over M_\phi}  \sqrt{1- 4 {M_{Z^\prime}^2 \over M_\phi^2}} \left( {M_\phi^4 \over 4 M_{Z^\prime}^4 } - {M_\phi^2 \over M_{Z^\prime}^2 } + 3\right)
\end{eqnarray} 
For  $M_{Z^\prime} < M_\phi < 2M_{Z^\prime}$, the decay channel turns to be $\phi \to Z^\prime Z^{\prime *} \to Z^\prime \bar f f$.  With neutrino pairs in the final state, the decay rate can be written as 
\begin{eqnarray}
\Gamma(\phi \to Z^\prime \bar \nu \nu) = { \alpha^2 M_\phi \over 24 \pi} F(x)
\end{eqnarray}
where $x=M_{Z^\prime}/M_\phi$ and ~\cite{Keung:1984hn} 
\begin{eqnarray}
F(x) &=& {3(1-8x^2 +20x^4) \over (4x^2 -1)^{1/2} } \arccos\left( {2 x^2-1\over 2 x^3}\right) -(1-x^2) \left({47\over 2 } x^2 -{13 \over 2} + {1\over x^2}\right)
\nonumber \\
&&-3(1-6x^2 + 4x^4 ) \ln(x) \; .
\end{eqnarray}
For $M_\phi <M_{Z^\prime} $, the possible decay channels of $\phi$ are to four leptons or six photons mediated by virtual $Z^\prime$. In this case,  the decay rate can only be calculated  numerically.  In short, $\phi$ cannot be stable unless it is ultra-light. 

As can be seen in Eq.(1), $\chi$ is vector-like fermion  with respect to the $U(1)_{L_\mu-L_\tau}$, so there is no interaction between $\chi$ and $\Phi$.  
However there can be a Yukawa interaction $Y_\chi^{} \overline{\ell_L^\mu} \tilde H\chi_R$, which may lead to an unstable $\chi$. We need to introduce a $Z_2$ symmetry, under which only $\chi$ is odd and all other particles are even, to forbid this Yukawa interaction.  As a result, $\chi$ is a stable DM candidate and only couples to the lepton sector via the gauge portal.   

\section{dark matter phenomenology}

\subsection{The relic density}

The DM $\chi$ can be thermalized with the thermal bath via the gauge interaction in the early Universe and its evolution is described by the Boltzmann equation,
\begin{eqnarray}
\frac{d n_\chi}{d t} + 3 H n_\chi = -\langle \sigma v \rangle \left( (n_\chi)^2 - (n^{\rm eq}_\chi)^2 \right),
\label{eq:boltzmann}
\end{eqnarray}
where $n_\chi$ is the number density of $\chi$, $H$ is the Hubble constant  and  $\langle \sigma v \rangle$ is the thermal average of the reduced annihilation cross section. The reduced annihilation cross section are 
\begin{eqnarray}
\sigma v( \bar \chi  \chi \to \bar \ell_\alpha  \ell_\alpha ) &=&{8\pi \alpha_X^2} \sqrt{1-{m_\alpha^2 \over m_\chi^2}}{2m_\chi^2 + m_\alpha^2 \over (4m_\chi^2 -M_{Z^\prime}^2)^2 + M_{Z^\prime}^2 \Gamma_{Z^\prime}^2}, \\
\sigma v( \bar \chi  \chi \to Z^\prime Z^\prime) &=&\frac{4\pi \alpha_X^2 (1- M^2_{Z^\prime}/m^2_\chi)^{3/2}}{m^2_\chi (2- M^2_{Z^\prime}/m^2_\chi)^2},
\end{eqnarray}
where $m_\alpha$ denotes lepton mass. 
For the annihilation into neutrinos, one needs to include  an extra factor of 1/2 in the cross section. Note that for $M_\phi+M_{Z^\prime}<2m_\chi$, an extra channel $\bar{\chi}\chi \to \phi  Z^{\prime}$ opens up and it complicates our physics picture, so we take the singlet scalar mass to be larger than $2m_\chi$ threshold for the simplicity. The thermal average of the reduced annihilation cross section can be written as $\langle\sigma v \rangle= a + b \langle v^2 \rangle + \mathcal{O}(v^4)$, where $a$ and $b$ are the s-wave and p-wave terms, respectively.

By solving the Eq. (\ref{eq:boltzmann}), the DM relic density is ~\cite{Kolb:1990vq,Jungman:1995df}, 
\begin{equation}
\Omega_{\mathrm{DM}} h^2= \frac{ 2.08\times10^9~\mathrm{GeV}^{-1} x_f}{M_{\mathrm{pl}}\sqrt{g_\ast(T_f)}(a+3b/x_f)} \label{relic-d},
\end{equation}
where $\langle \sigma_{\rm ann} v \rangle= a+b/x_f$ and $x_f\equiv m_\chi/T_f$ with $T_f$ being the freeze-out temperature, $g_\ast(T_f)$ is the total number of effective relativistic degrees of freedom when the DM freezes out, $M_{\mathrm{pl}}$ is the Planck mass.
 The parameter $x_f$ is given by,
\begin{equation}
x_f=\ln\left[c(c + 2)\sqrt{\frac{45}{8}}
\frac{gM_\chi M_{\mathrm{pl}}(a+6b/x_f)}{2\pi^3\sqrt{g_\ast(x_f)}\,x_f^{1/2}}\right]\label{x-f},
\end{equation}
where $c$ is a constant of order one.

\begin{figure}
\centering
\includegraphics[width=0.48\textwidth]{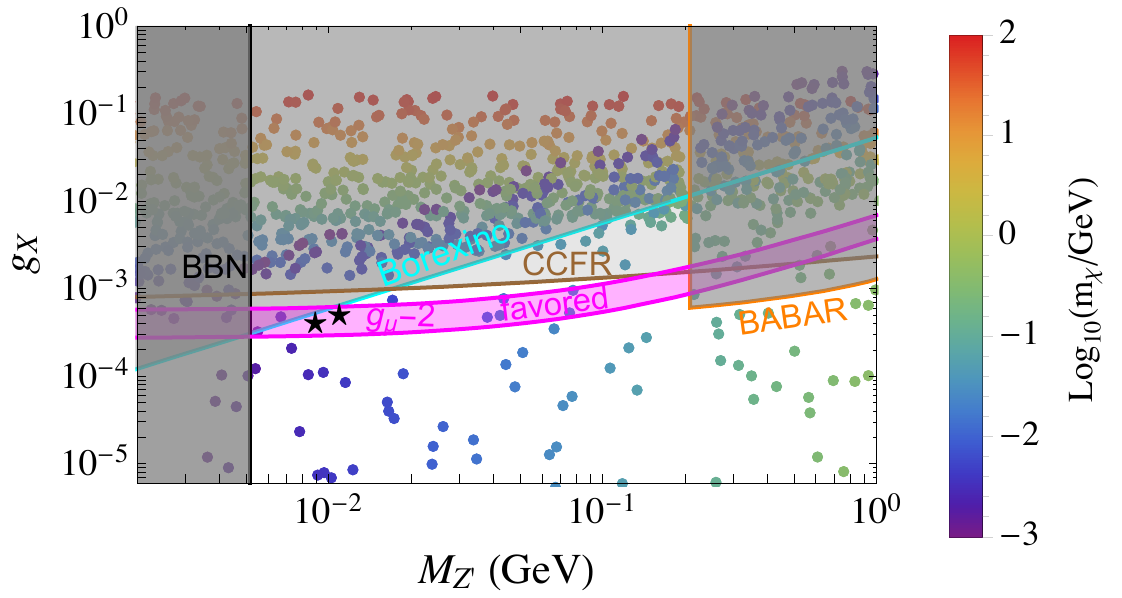}
\hspace{0.1cm}
\includegraphics[width=0.48\textwidth]{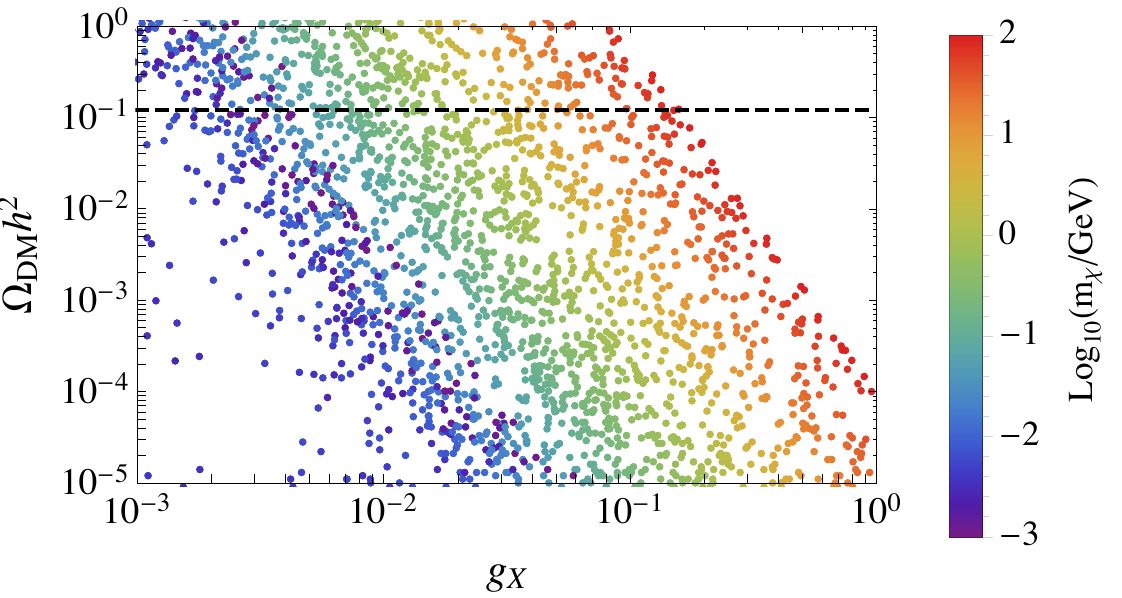}
\caption{The left panel shows the allowed points in $g_X-M_{Z^\prime}$ plane that may give rise to the observed relic density $\Omega_{\rm DM} h^2=0.12$. The right panel illustrates the correlation between $g_X$ and $\Omega_{\rm DM} h^2$ as a function of $m_\chi$ for $M_{Z^{\prime}}=10^{-2}$ GeV.
} \label{fig:mchi_mzp}
\end{figure}

In this work, we use \textbf{ Feynrules}~\cite{Alloul:2013bka} to obtain the model files for the \textbf{Calchep}~\cite{Belyaev:2012qa} and also  use the \textbf{MicrOMEGAs}~\cite{Belanger:2013oya} to calculate the DM relic density as well as the reduced annihilation cross section. In Fig.~\ref{fig:mchi_mzp}, we show constraints on the parameter spaces of the model by several physical observables. 
The plot on the left panel shows the allowed parameter space in the $(M_{Z^\prime}, g_X)$ plane  that may give rise to a correct relic density $\Omega_{\rm DM} h^2=0.12$ (scattering points). The gray shadowed regions are already excluded. Among these regions, the black, cyan, brown and orange lines denote constraints from BBN~\cite{Araki:2015mya, Kamada:2015era}, Borexino~\cite{Araki:2015mya}, CCFR~\cite{Altmannshofer:2014pba} and BABAR~\cite{TheBABAR:2016rlg}, respectively. 
The magenta band is the favored region of the muon $g-2$~\cite{Bennett:2006fi}. The two black stars denote the reference points to explain IceCube results~\cite{Araki:2015mya}. 
%
%
%
For points in the allowed parameter space, we have $m_{Z^\prime} \sim 2 m_\chi^{} $, which results in a relatively small $g_X$ as the annihilation cross section is resonantly enhanced.
%
%
On the right-panel of the Fig.~\ref{fig:mchi_mzp} we show the scattering plot of $\Omega h^2 $ as the function of  the coupling $g_X$ by setting $m_{Z^\prime}=10^{-2}~{\rm GeV}$. The black dashed line denotes the observed dark matter relic density.
It shows  the relic density is inversely proportional to $g_X^4$ and increases as the increase of  $m_\chi^{}$.
%

\subsection{The dark matter scattering off electron}

\begin{figure}
\centering
\includegraphics[width=0.49\textwidth]{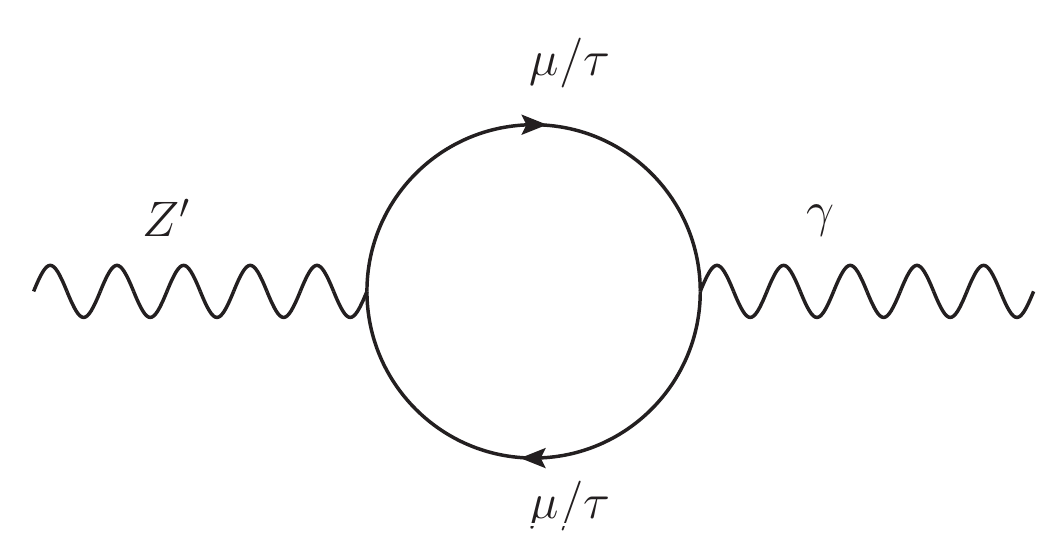}
\caption{The kinetic mixing between $Z^\prime$ and photon at the one loop with virtual $\mu$ ant $\tau$ leptons~\cite{Gninenko:2018tlp}.
} \label{fig:mixing}
\end{figure}

Although  the kinetic mixing between $Z^\prime$ and photon is absent at the tree level, the mixing can be generated at the one-loop level by virtual $\mu$ ant $\tau$ leptons~\cite{Gninenko:2018tlp}, as illustrated in the Fig.~\ref{fig:mixing},
\be
\epsilon=\frac{eg_X}{6\pi^2}{\rm ln}\left( \frac{m_\mu}{m_\tau}\right).
\ee
The cross section for the DM scattering off the electron when the momentum transfer is much smaller than the mediator mass $M_{Z^{\prime}}$ can be written as,
\be
\sigma_{\chi e}=\frac{16\pi \epsilon^2 \alpha \alpha_X \mu^2_{\chi e}}{M^4_{Z^{\prime}}},
\label{eq:sigma_xenon1t}
\ee
where $\mu_{\chi e}$ is the reduced mass of dark matter and electron.

\begin{figure}
\centering
\includegraphics[width=0.49\textwidth]{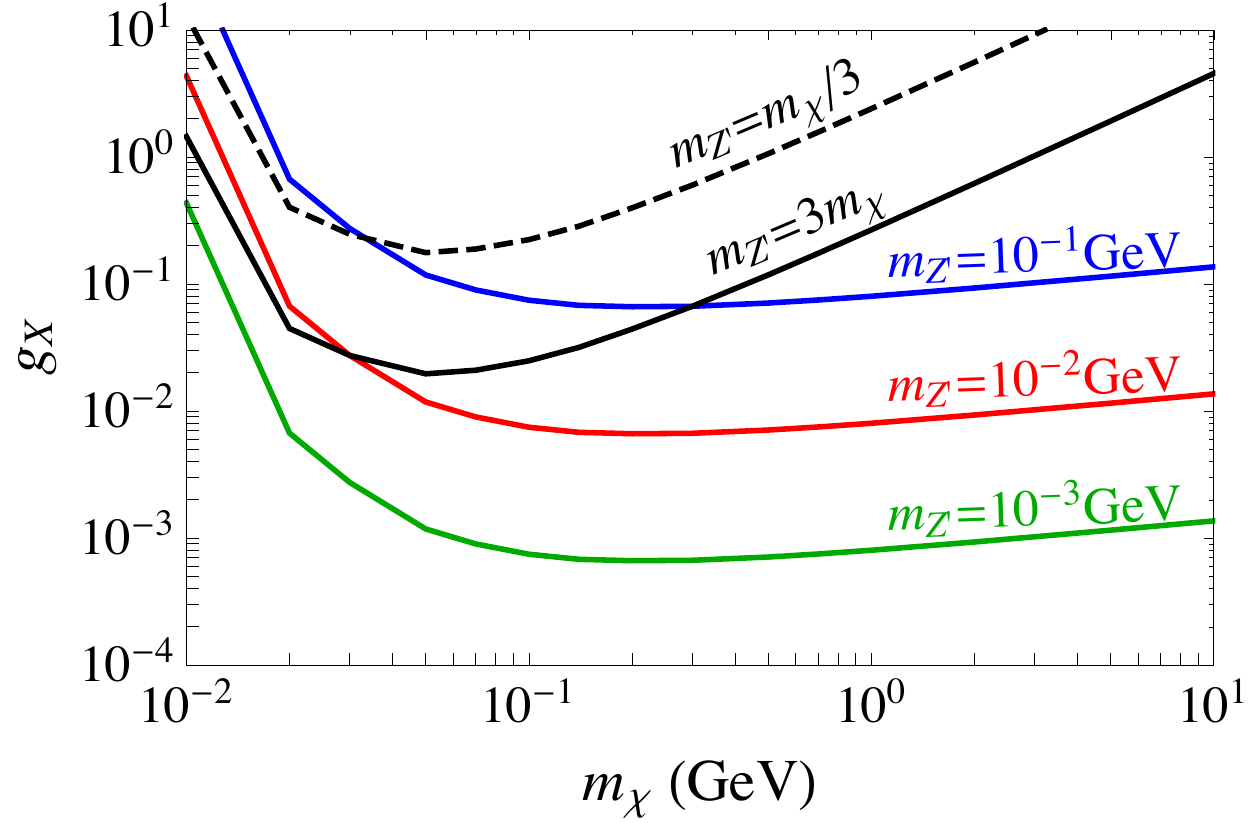}
\caption{The XENON1T~\cite{Aprile:2019xxb} constraints for DM-electron scattering cross section with  fixed values of $M_{Z^{\prime}}=10^{-1}~{\rm GeV}~(\rm blue), 10^{-2}~{\rm GeV}~(\rm red), 10^{-3}~{\rm GeV}~(\rm green)$  or with $M_{Z^{\prime}}=m_\chi/3$ (black solid line), $3m_\chi$(black dashed lines).
} \label{fig:mchi_gx}
\end{figure}

In Fig.~\ref{fig:mchi_gx}, we show in the $m_\chi-g_X$ plane the exclusion limits given by the XENON1T~\cite{Aprile:2019xxb} for DM-electron scattering cross section, where the blue, red and green solid lines as well as black solid and dashed lines correspond to $M_{Z^{\prime}}=10^{-1}~{\rm GeV}, 10^{-2}~{\rm GeV}, 10^{-3}$ GeV and  $M_{Z^{\prime}}=m_\chi/3$, $3m_\chi$, respectively.  
This constraint, together with these from low energy precision measurements, puts upper bounds on the new gauge coupling.
%

\section{neutrino oscillations}
Neutrino oscillation opens an important window for probing new physics beyond the SM. The neutrino-medium interaction  can significantly change the behavior of neutrino oscillations~\cite{Wolfenstein:1977ue,Mikheev:1986gs}. In addition to the SM charged current and neutral current interactions, there can be other non-standard neutrino interactions which can modify the propagation of neutrinos and thus alter the neutrino oscillation probabilities. For useful reviews see e.g. Refs.\cite{Antusch:2008tz,Ohlsson:2012kf,Farzan:2017xzy} and references cited therein.
In the $L_\mu-L_\tau$ model,  the DM-neutrino interaction may  induce extra matter effect in neutrino oscillations. 
We focus on the asymptotic behavior of neutrinos when DM density is large and the ordinary matter effect (i.e., electrons, protons and neutrons) can be ignored, which is similar to the case of  dense matter effect~\cite{Xing:2018lob,Huang:2018ufu,Luo:2019efb}.


When neutrinos propagate in the DM, their evolution equation is modified by the effective potential due to the interactions with the DM through coherent forward elastic scatterings.  The effective potential for muon and tau neutrinos are 
\begin{eqnarray}
V_\chi = \pm {g_X^2 \over m_{Z^\prime}^2 } n_\chi \; , \label{effectivep}
\end{eqnarray}
with the positive (negative) sign for muon (tau) neutrino, where $n_\chi$ is the DM density. The evolution equation of the flavor transition amplitude is
\begin{eqnarray}
i{d \over d x} \Psi_\alpha =\tilde{\cal H} \Psi_\alpha ,
\end{eqnarray}
where $\tilde{\cal H}$ is the effective Hamiltonian in DM. Given the effective potential in Eq. (\ref{effectivep}), we can write down the effective Hamiltonian $\tilde{\cal{H}}$, which differs from the Hamiltonian in vacuum ${\cal H}$, in the flavor basis
\begin{eqnarray}
\tilde{\cal H} & = & {\cal H} + {\cal H}' \; = \; \frac{1}{2 E} \left [ V \left ( \begin{matrix} m^{2}_{1} & & \\ & m^{2}_{2} & \\ & & m^{2}_{3}  \end{matrix} \right ) V^{\dagger}_{} + \left ( \begin{matrix} A_{CC} & & \\ & A_\chi^{} & \\ & & -A_\chi^{}  \end{matrix} \right ) \right ] \; ,
\end{eqnarray}
where $A_\chi^{}=2EV_\chi$, $A_{CC}=2EV_{CC}$ with $V_{CC}$ the effective potential coming from the SM charge current interaction. For neutrinos $A_\chi>0$ and for anti-neutrinos $A_\chi<0$. $V$ is the $3 \times 3$ unitary Pontecorvo-Maki-Nakagawa-Sakata (PMNS) lepton mixing matrix~\cite{Maki:1962mu,Pontecorvo:1967fh}, in which ${ \theta_{12}, \theta_{13}, \theta_{23} }$ are three mixing angles and $\delta$ is the Dirac CP phase~\cite{Tanabashi:2018oca},
\begin{eqnarray}
V \; = \; \left ( \begin{matrix} c^{}_{12} c^{}_{13} & s^{}_{12} c^{}_{13} & s^{}_{13} e^{-i\delta}_{} \\ -s^{}_{12} c^{}_{23} - c^{}_{12} s^{}_{23} s^{}_{13} e^{i\delta}_{} & c^{}_{12} c^{}_{23} - s^{}_{12} s^{}_{23} s^{}_{13} e^{i\delta}_{} & s^{}_{23} c^{}_{13} \\ s^{}_{12} s^{}_{23} - c^{}_{12} c^{}_{23} s^{}_{13} e^{i\delta}_{} & - c^{}_{12} s^{}_{23} - s^{}_{12} c^{}_{23} s^{}_{13} e^{i\delta}_{} & c^{}_{23} c^{}_{13} \end{matrix} \right ) \; ,
\end{eqnarray}
with $c^{}_{ij} \equiv \cos\theta^{}_{ij}$ and $s^{}_{ij} \equiv \sin\theta^{}_{ij}$ (for $ij = 12, 13, 23$). In our case, the effect of  the charged current interaction is ignored ($A_{CC}=0$) for simplicity. 
Note that the term proportional to the identity matrix does not affect the neutrino oscillation behaviors, so we can ignore the $m^{2}_{1}$ term, the Hamiltonian $\tilde{\cal{H}}$ can be rewritten  as,
\begin{align}
\tilde{\cal{H}} = \frac{1}{2 E} \left [  V \left ( \begin{matrix}  0  & & \\ & \Delta m^{2}_{21} & \\ & & \Delta m^{2}_{31}  \end{matrix} \right ) V^{\dagger} + \left ( \begin{matrix} 0 & & \\ & A_\chi^{} & \\ & &  -A_\chi^{}  \end{matrix} \right ) \right ] \; 
= \frac{1}{2 E} \tilde{V} \left ( \begin{matrix} \tilde{m}^{2}_{1} & & \\ & \tilde{m}^{2}_{2} & \\ & & \tilde{m}^{2}_{3} \end{matrix} \right ) \tilde{V}^{\dagger}_{} \; , \label{eigen}
\end{align}
where the $\tilde{V}$ is the mixing matrix in DM and the $\tilde m^{2}_{i}$($i=1,2,3$) are the  eigenvalues. One can obtain  $\Delta\tilde m^{2}_{ji}=\tilde{m}^{2}_{j} - \tilde{m}^{2}_{i}$ by solving Eq.~(\ref{eigen}) numerically. The best-fit values of the neutrino oscillation parameters in vacuum are
summarized in the Table~\ref{tab1} , which are adopted as inputs in following numerical calculations. 

\begin{table}
	\centering
	\begin{tabular}{ccc}
		\hline
		\hline
		& Normal Mass Ordering & Inverted Mass Ordering\\ \hline
		$\theta_{12}$& 33.82$^{\circ}$ &33.82$^{\circ}$ \\ 
		$\theta_{13}$ &8.61$^{\circ}$ &8.65$^{\circ}$ \\
		$\theta_{23}$ & 49.7$^{\circ}$& 49.7$^{\circ}$\\
		$\delta$ & 217$^{\circ}$& 280$^{\circ}$\\
		$\Delta m^2_{21}[10^{-5}~{\rm eV^2}]$&7.39&7.39 \\
		$\Delta m^2_{31}[10^{-3}~{\rm eV^2}]$&2.451&-2.512\\ 
		\hline
		\hline
	\end{tabular}
	\caption{Three-flavor Oscillation parameters from a global fit data of current experimental data \cite{Esteban:2018azc}. Note that  $\Delta m^2_{31}<0 $ for (IO).}
	\label{tab1}
\end{table}

\begin{figure}[tbp]
\centering 
\includegraphics[width=0.48\textwidth]{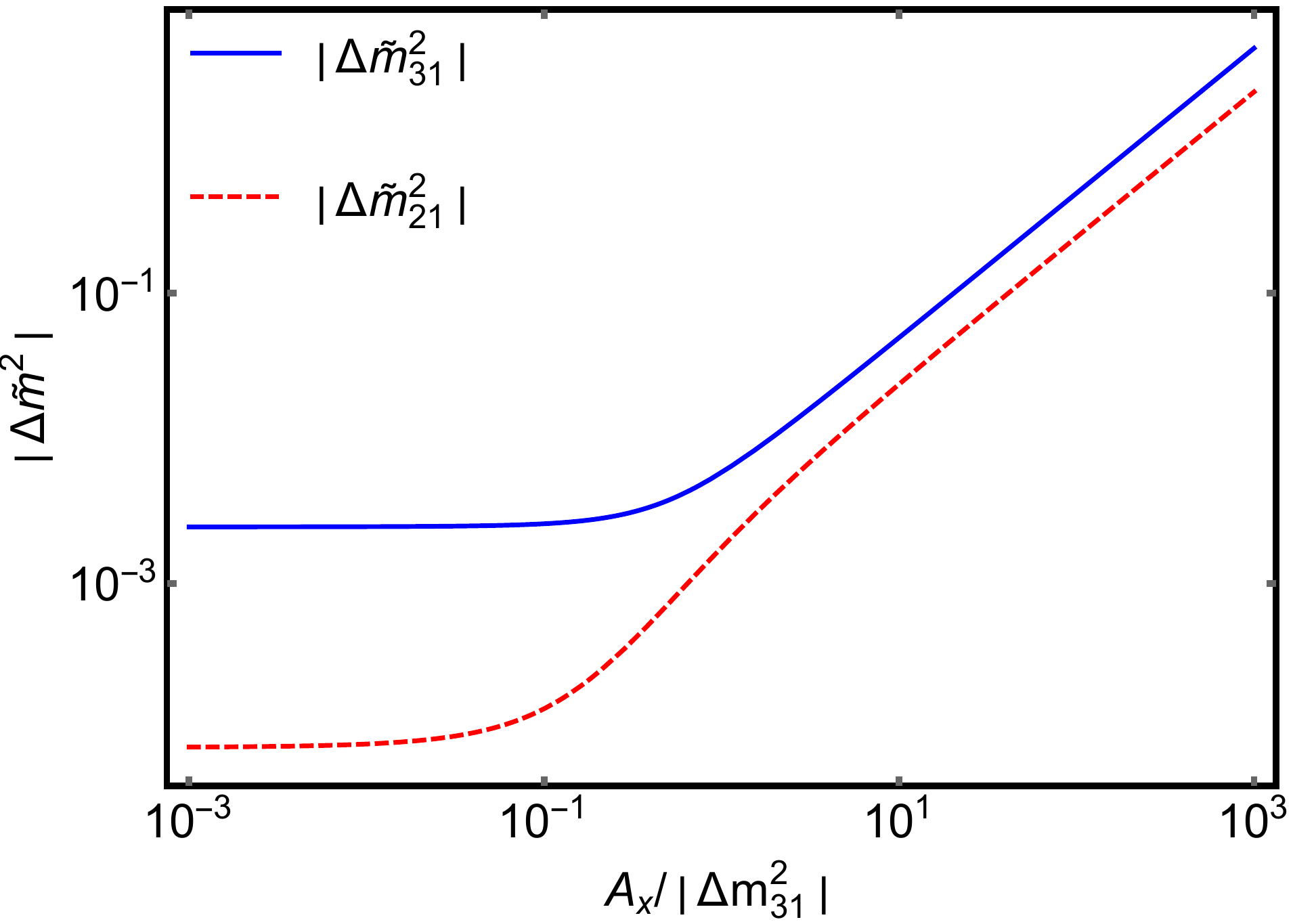}
\hfill
\includegraphics[width=0.48\textwidth]{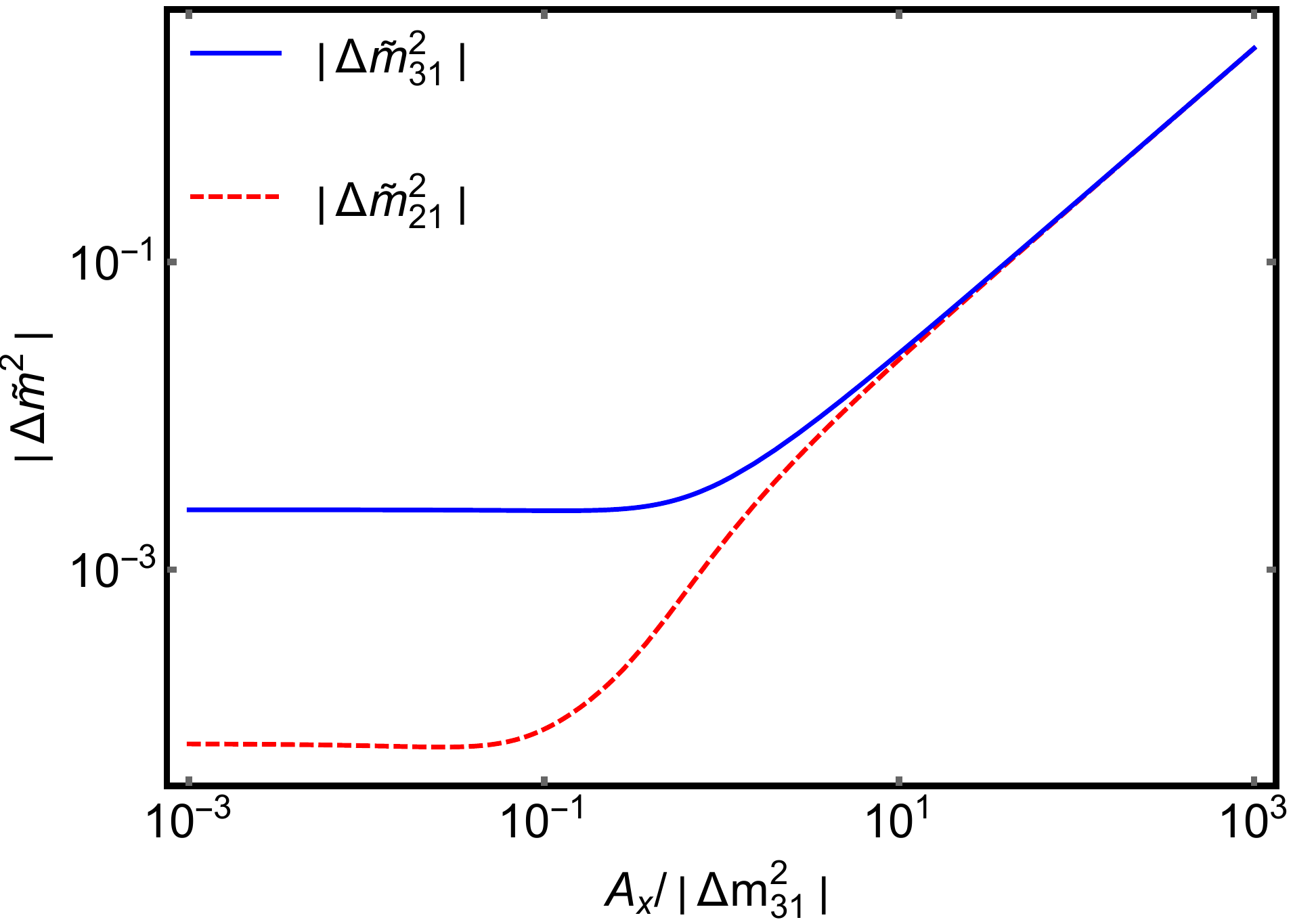}
\caption{\label{eigenvalue} The modulus of the neutrino mass-squared differences in matter $|\Delta\tilde m^2_{21}|$(red and dashed) and $|\Delta\tilde m^2_{31}|$(blue and solid)  with respect to ${A_{\chi}}/|{\Delta m^2_{31}}|$ in the normal and inverted mass hierarchies, respectively.}
\end{figure}

We show the modulus of the mass-squared differences $|\Delta \tilde m^2_{21}|$ and $|\Delta \tilde m^2_{31}|$ as the function of $A_\chi/|\Delta m_{31}^2|$  in the Fig.~\ref{eigenvalue}, where plots in the left-panel and right-panel correspond to the normal and inverted mass hierarchies, respectively. It is helpful to discuss the asymptotic behavior of them. 
When $|A_\chi|\rightarrow 0$, neutrinos are propagating in the ``vacuum-dominated" region, and there is almost no DM effect in neutrino oscillations. 
The DM effect becomes significant for $|A_{\chi}|\sim |\Delta m^2_{31}|$. 
While $|A_\chi|\rightarrow\infty$, in other words,  $|A_{\chi}| \gg |\Delta m^2_{31}|$, the neutrinos are propagating in the ``DM-dominated" region. Three eigenvalues of $\tilde{\cal{H}}$ are separated and neutrino oscillation can hardly happen in this region.

\begin{figure}[tbp]
\centering 
\includegraphics[width=0.95\textwidth]{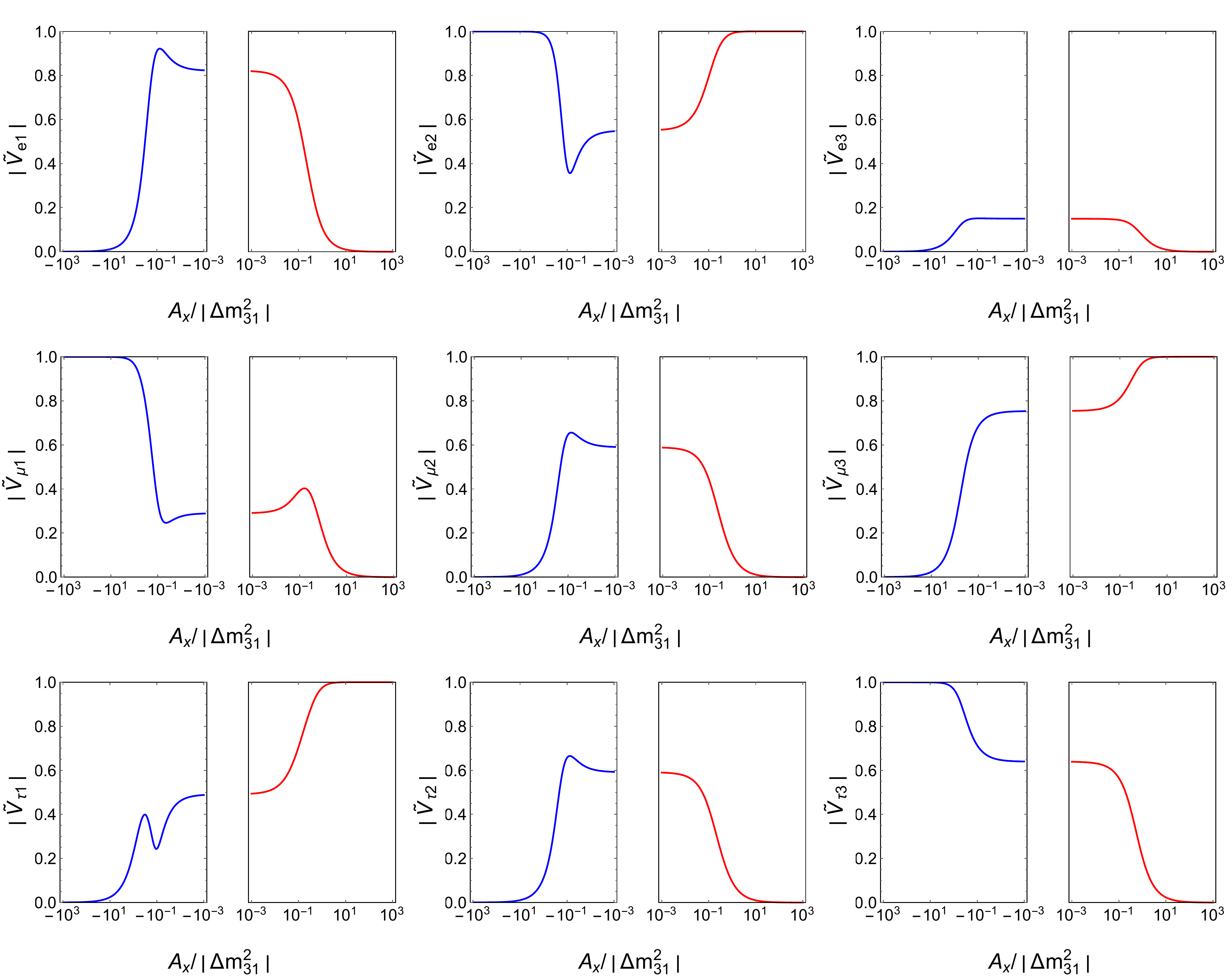}
\caption{\label{nv} 
The evolutions of nine effective mixing matrix elements $|\tilde V_{\alpha i}|$ (for $\alpha=e, \mu, \tau$, and $i=1, 2, 3$ ) with respect to the ${A_{\chi}}/{|\Delta m^2_{31}|}$ in the normal  hierarchy for both neutrinos(red, right lines) and anti-neutrinos(blue, left lines).
}
\end{figure}

\begin{figure}[tbp]
\centering 
\includegraphics[width=0.95\textwidth]{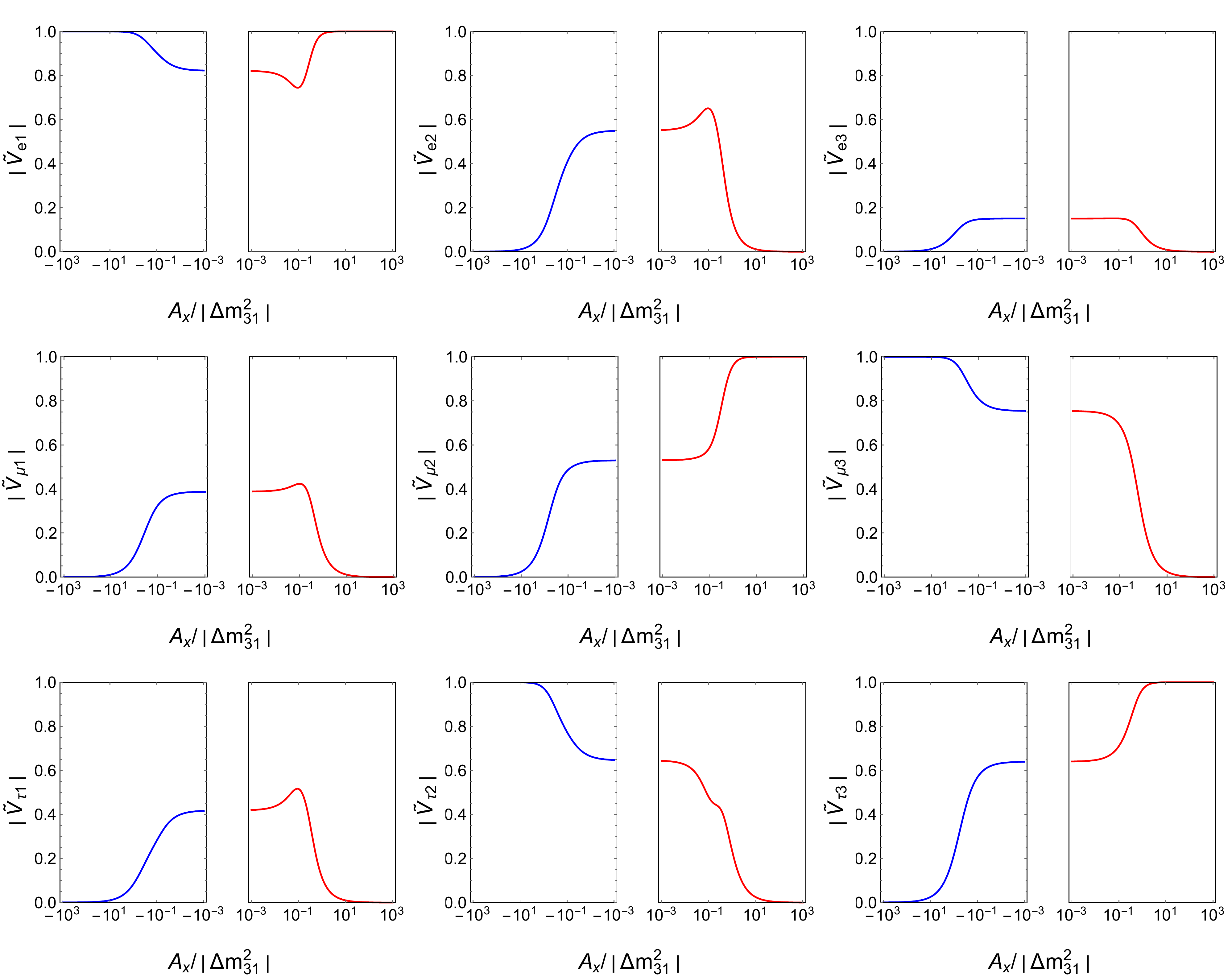}
\caption{\label{iv} 
The evolution behaviors of nine effective mixing matrix elements $|\tilde V_{\alpha i}|$ (for $\alpha=e, \mu, \tau$, and $i=1, 2, 3$ ) with respect to the ${A_{\chi}}/{|\Delta m^2_{31}|}$ in the inverted hierarchy for both neutrinos(red, right lines) and anti-neutrinos(blue, left lines).
}
\end{figure}

In the Fig.~\ref{nv} and the Fig.~\ref{iv},  we illustrate the evolution behaviors of the matrix elements $|\tilde V_{\alpha i}|$,  which are derived from the eigenvector-eigenvalue identity~\cite{Denton:2019pka},
\begin{eqnarray}
|\tilde V_{\alpha i}|^2=\frac{(\lambda_i-\xi_\alpha)(\lambda_i-\zeta_\alpha)}{(\lambda_i-\lambda_j)(\lambda_i-\lambda_k)}=\frac{\lambda_i^2-\lambda_i(\xi_\alpha+\zeta_\alpha)+\xi_\alpha\zeta_\alpha}{(\lambda_i-\lambda_j)(\lambda_i-\lambda_k)}\,,  \label{tildeV}
\end{eqnarray}
where $\alpha\in\{e,\mu,\tau\}$, $i, j, k\in \{1,2,3\}$ and $i\neq j \neq k$.  $\lambda_i/2E$ is the eigenvalue of $\tilde {\cal H}$. $\xi_\alpha/2E$ and $\zeta_\alpha/2E$ are  the eigenvalues of the $2\times2$ submatrix $\tilde{\cal{H}}_\alpha$,
\begin{eqnarray}
\tilde{\cal{H}}_\alpha\equiv
\begin{pmatrix}
\tilde{\cal{H}}_{\beta\beta}&\tilde{\cal{H}}_{\beta\gamma}\\
\tilde{\cal{H}}_{\gamma\beta}&\tilde{\cal{H}}_{\gamma\gamma}
\end{pmatrix}\,,  \label{sub}
\end{eqnarray}
which is the residual matrix of $\tilde H$ after removing the row $\alpha$ and the column $\alpha$. It is easy to prove that~\cite{Denton:2019pka},
\begin{eqnarray}
\xi_\alpha+\zeta_\alpha&=&(2E)(\tilde{\cal{H}}_{\beta\beta}+\tilde{\cal{H}}_{\gamma\gamma}) \label{deltaa}\\
\xi_\alpha\zeta_\alpha&=&(2E)^2(\tilde{\cal{H}}_{\beta\beta}\tilde{\cal{H}}_{\gamma\gamma}-\tilde{\cal{H}}_{\beta\gamma}\tilde{\cal{H}}_{\gamma\beta}) \label{deltab} \; .
\end{eqnarray}
By substituting Eqs.~(\ref{deltaa}) and (\ref{deltab}) into Eq.(\ref{tildeV}), we can get numerical values of the nine lepton mixing matrix elements in DM.  We can conclude from the Fig.~\ref{nv} and~\ref{iv} that, corrections to $|\tilde V_{\alpha i}|$ are very small for  $|A_{\chi}|\ll |\Delta m^2_{31}|$. $|\tilde V_{\alpha i}|$ receive dramatic corrections from the DM when $|A_{\chi}|\sim |\Delta m^2_{31}|$. 
For $|A_\chi|\rightarrow\infty$,  the neutrino flavors decouple from each other. 

Mixing angles in DM can be derived numerically from the PMNS matrix as
\begin{eqnarray}
\tilde s_{12}&=&\frac{|\tilde V_{e2}|}{\sqrt{1-|\tilde V_{e3}|^2}} \; , \\
\tilde s_{13}&=&|\tilde V_{e3}| \; , \\
\tilde s_{23}&=&\frac{|\tilde V_{\mu 3}|}{\sqrt{1-|\tilde V_{e3}|^2}}, 
\end{eqnarray}
where $|\tilde V_{\alpha i}|$ are given in the Eq.(\ref{tildeV}).

To get the Jarlskog invariant $\tilde{\cal J}$ in DM,  which is defined as ${\cal J} \equiv {\rm Im} \left ( V_{\alpha i} V^{*}_{\beta i} V^{*}_{\alpha j} V_{\beta j} \right )$ $\times$ $ \sum^{}_{\gamma, k} \epsilon_{\alpha \beta \gamma} \epsilon_{ijk}$ (for $\alpha, \beta, \gamma = e, \mu, \tau$ and $i, j, k = 1, 2, 3$)~\cite{Jarlskog:1985ht, Wu:1985ea}, 
we use the identity in~\cite{Xing:2000ik},
\begin{eqnarray}
\tilde{\cal J}{\Delta}\tilde m^{2}_{21}{\Delta} \tilde m^{2}_{31}{\Delta}\tilde m^{2}_{32}={\cal J}\Delta m^{2}_{21}\Delta m^{2}_{31}\Delta m^{2}_{32} \label{Jarl},
\end{eqnarray}
where $\tilde{\cal J}$ is the Jarlskog in DM and $ {\cal J}$ is the Jarlskog in vacuum.   The same relationship has been applied to study neutrino oscillations in ordinary matter~\cite{Naumov:1991ju,Harrison:1999df}. The Dirac CP phase $\delta$ can be extracted by inserting the explicit expression of Jarlskog,
\begin{eqnarray}
{\cal J}&=&{\rm Im} (V_{e1} V^{*}_{\mu 1} V^{*}_{e2} V_{\mu 2})=s_{23}c_{23}s_{13}c^2_{13}s_{12}c_{12}\sin\delta
\end{eqnarray}
into the Eq.~(\ref{Jarl}) .

\begin{figure}[tbp]
\centering 
\includegraphics[width=0.48\textwidth]{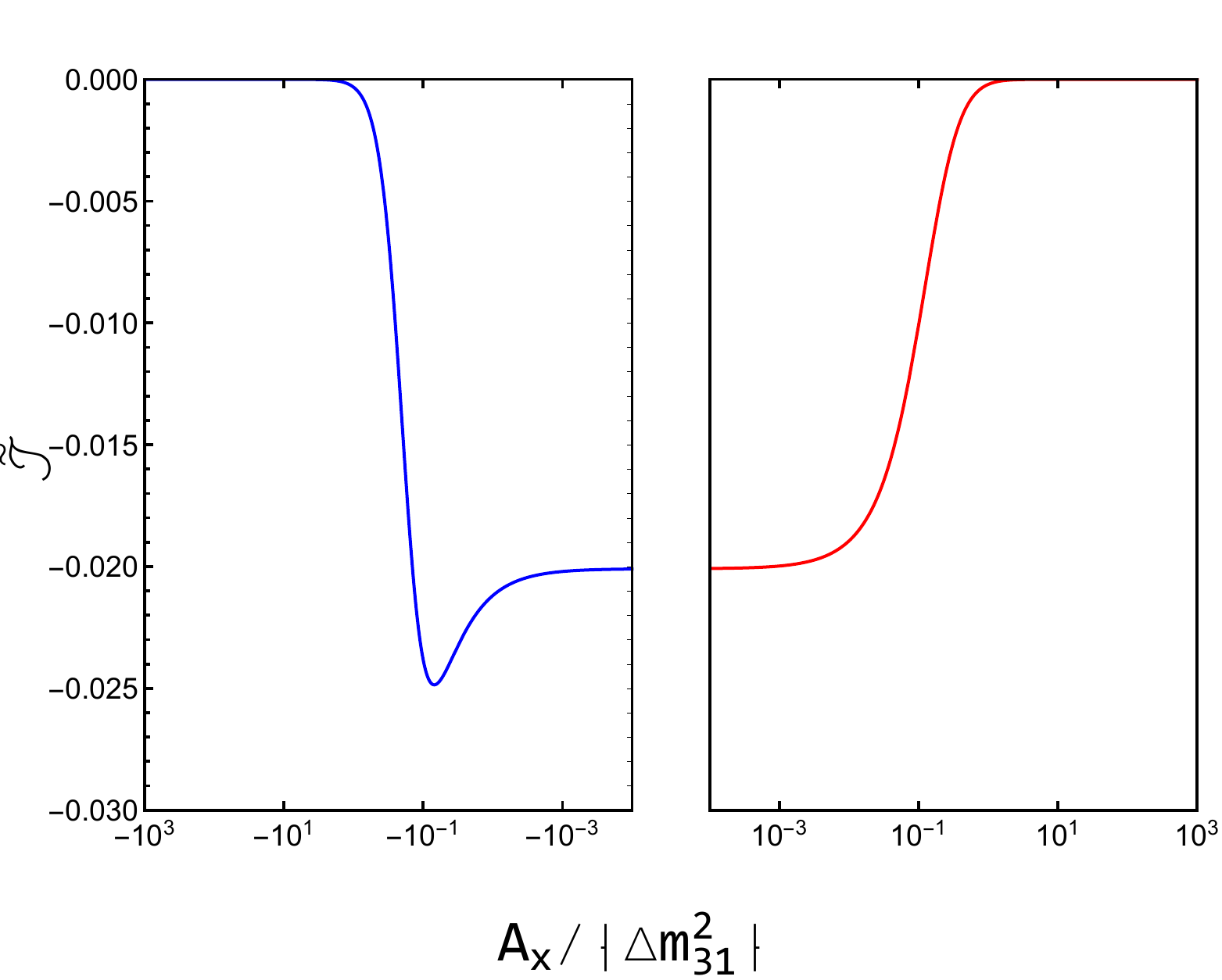}
\hfill
\includegraphics[width=0.48\textwidth]{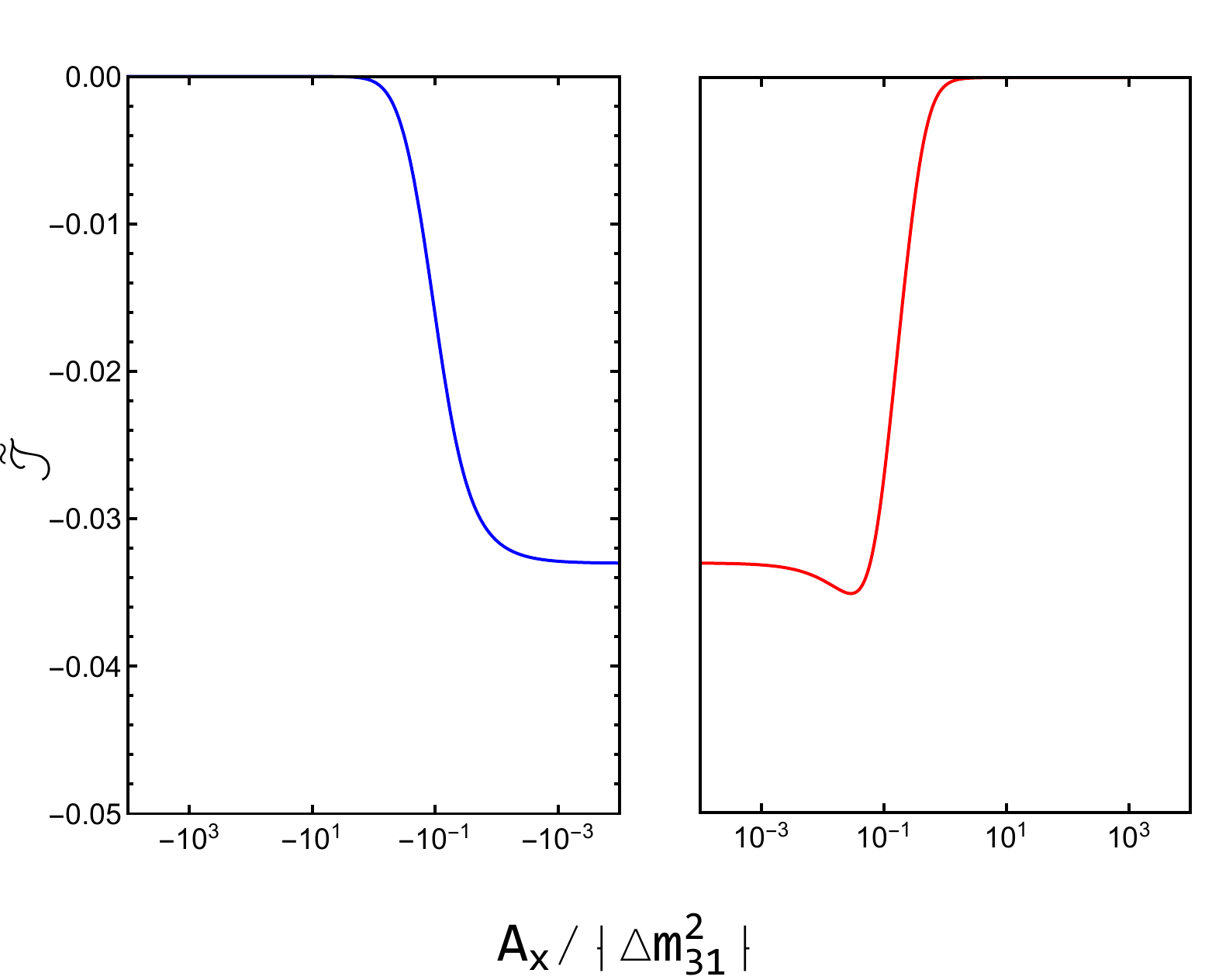}
\caption{\label{mj} 
The evolution of the Jarlskog invariant $\tilde{\cal J}$ in DM with respect to the ${A_{\chi}}/{|\Delta m^2_{31}|}$ in the normal mass ordering and inverted mass ordering for neutrinos(red curves in the right half panel) and anti-neutrinos(blue curves in the left half panel). In the limit $|A_{\chi}| \rightarrow \infty$, $\tilde{\cal J}$ approaches to zero. 
}
\end{figure}

We show the effective Jarlskog invariant  in DM $\tilde{\cal J} $ as the function of ${A_\chi}/{|\Delta m^{2}_{31}|}$ in the Fig.\ref{mj}  for normal hierarchy (left-panel)  and inverted hierarchy (right-panel), respectively. We can see that, $\tilde{\cal J}$ approaches to zero for $|A_{\chi}| \gg |\Delta{m^2_{31}}|$, this is because the mixing angle $\tilde s_{13}$  tends to zero in this case.

Given all the elements $\tilde s_{ij}$, $\tilde c_{ij}$ and the Dirac CP phase $\tilde\delta$, we can easily write down the neutrino oscillation probabilities in DM, which is the same as the oscillation probabilities in vacuum up to replacements $V_{\alpha i} \to \tilde{V}_{\alpha i}$, $\Delta m_{ji}^2 \to \Delta \tilde m_{ji}^2$ and $\delta \to \tilde \delta$,
\begin{eqnarray}
\tilde P( \stackrel{(-)}{\nu}^{}_{\alpha} \rightarrow  \stackrel{(-)}{\nu}^{}_{\beta} )&=& \delta^{}_{\alpha \beta} - 4 \sum^{}_{j > i} {\rm Re} \left [ \tilde V_{\alpha i} \tilde V^{*}_{\beta i} \tilde V^{*}_{\alpha j} \tilde V_{\beta j} \right ] \sin^2 \tilde{\Delta}^{}_{ji} \nonumber \\ & &\pm 2 \sum^{}_{j > i} {\rm Im} \left [ \tilde V_{\alpha i} \tilde V^{*}_{\beta i} \tilde V^{*}_{\alpha j} \tilde V_{\beta j} \right ] \sin2\tilde{\Delta}^{}_{ji} ,  
\end{eqnarray}
where $\tilde{\Delta}^{}_{ji} \equiv \Delta \tilde{m}^2_{ji} L / 4E$ with $\Delta \tilde{m}^2_{ji} \equiv \tilde{m}^{2}_{j} - \tilde{m}^{2}_{i}$ being the effective neutrino mass-squared difference in DM, the Greek letters $\alpha$, $\beta$ are the flavor indices run over $e$, $\mu$, $\tau$, while the Latin letters $i$, $j$ are the indices of mass eigenstates run over $1$, $2$, $3$. $E$ is the energy of the neutrino/anti-neutrino beam.

\begin{figure}[tbp]
\centering 
\includegraphics[width=0.95\textwidth]{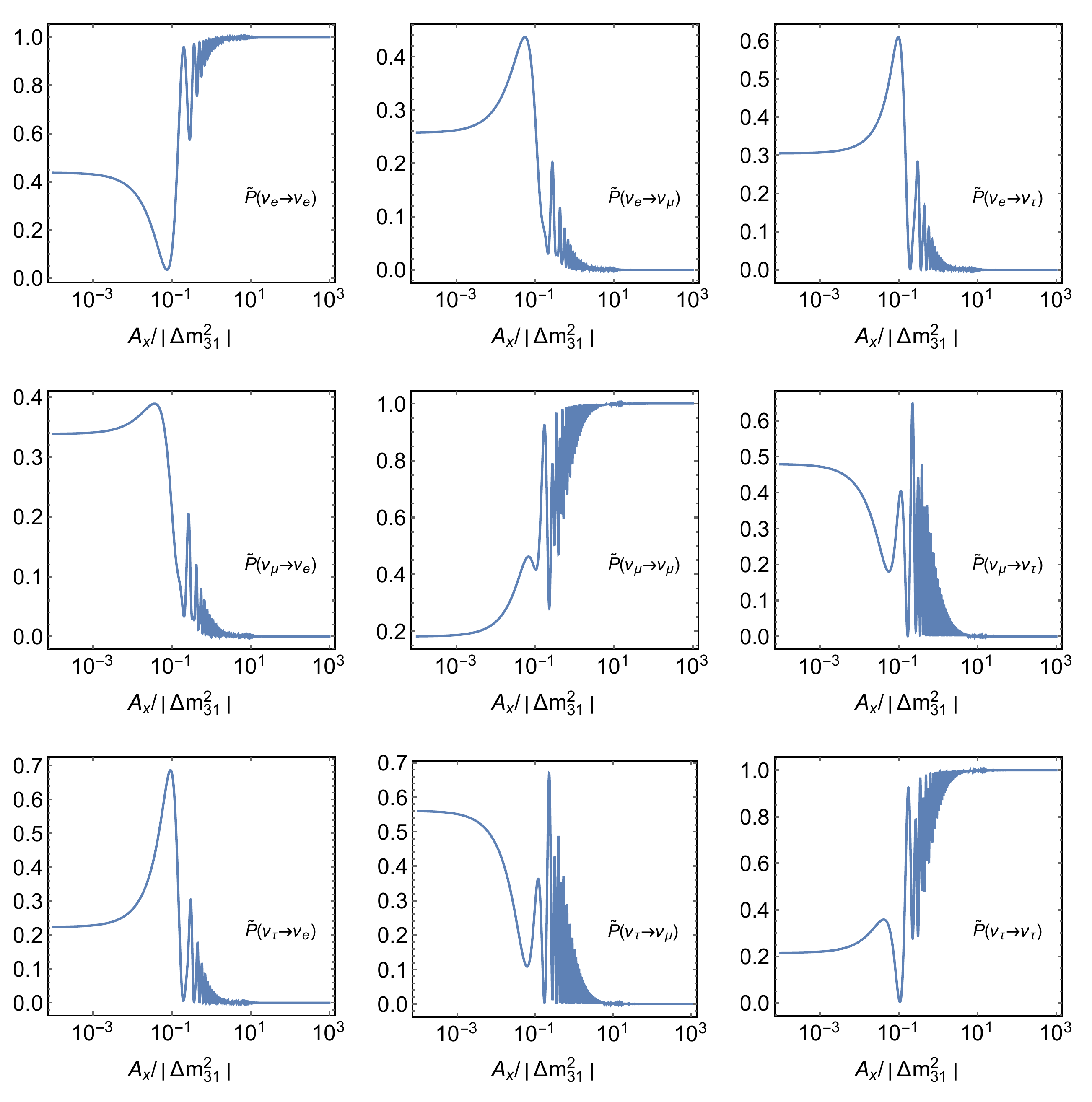}
\caption{\label{P} 
Neutrino oscillation probabilities in DM. In the limit $|A_\chi|\rightarrow \infty$, $\tilde P(\nu_e \rightarrow \nu_e),\tilde P(\nu_\mu \rightarrow \nu_\mu)$ and $\tilde P(\nu_\tau \rightarrow \nu_\tau)$ approach to 1, other channels approach to zero, which means that neutrino is decoherence in this case.
}
\end{figure}

As illustrations, we show all nine neutrino oscillation probabilities in DM  as the function of $A_{\chi} /|\Delta{m^2_{31}}|$ for the normal mass hierarchy in the Fig.~\ref{P}. 
We take the parameters listed in the Table.~\ref{tab1} as inputs and set $L/E=10^4$[km/GeV]  when making the plot.  We can conclude that the DM effect in neutrino oscillations can be significant for a sizable $A_{\chi}$, but neutrino oscillation may decouple for an ultra-large $A_\chi$.   Note that $n_\chi \sim 0.4~{\rm GeV/cm^3}$ on the Earth and ${\cal O}(g_\chi)\sim 10^{-3}$, it is unlikely to get a large $A_{\chi}$ except for a super high energy neutrino beam.  However  the DM density can be large in some sub-halo and DM stars can be formed in some asymmetric DM cases, which may result in a large $A_{\chi}$. Neutrino oscillations in these regions may provide indirect tests to the DM density.

\section{Discussions}

If dark matter couples to active neutrinos, neutrino properties  will be affected by additional matter effects  when they are travelling in the DM halo.   In this paper, we have introduced an additional neutral current interaction between neutrinos and DM by extending the SM with gauged $L_\mu-L_\tau$ symmetry, which introduce an extra effective potential to the evolution equation of the neutrino flavor transition amplitude. We showed that the high energy neutrino oscillations may undergo a matter dominated stage in a dense DM environment, where the neutrino masses, mixing angles as well as neutrino oscillation probabilities are very different compared with oscillation in vacuum.  Although it is  unable to test the matter effect induced by the DM in long baseline neutrino oscillation experiments as the DM density is too low on the Earth, our results can be applied to evaluate the DM density distribution indirectly by analyzing neutrino oscillation data of known astrophysical sourced neutrino beams.

\acknowledgments

This work was supported by the National Natural Science Foundation of China under grant No. 11775025 and the Fundamental Research Funds for the Central Universities under grant No. 2017NT17.

\bibliographystyle{JHEP}
\bibliography{reference.bib}

\end{document}